\documentclass[11pt, a4paper]{article}
\usepackage{moriond,epsfig}

\def\be{\begin{equation}}
\def\ee{\end{equation}}
\def\bea{\begin{eqnarray}}
\def\eea{\end{eqnarray}}

\begin{document}
\vspace*{4cm}
\title{STAR FORMATION IN GALAXY MERGERS: \\
Scaling up a universal process or a violent mode of SF?}

\author{ UTA FRITZE -- V. ALVENSLEBEN}
\address{Universit\"atssternwarte G\"ottingen, Geismarlandstr. 11, 
37083 G\"ottingen, Germany}

\maketitle\abstracts{I briefly review some measures of star formation rates in 
galaxies and discuss their respective uncertainties before outlining the range 
of star formation rates encountered in starbursts from isolated dwarf through 
massive gas-rich interacting systems. I present our current understanding of 
molecular cloud masses and structures and on star formation processes and 
efficiencies in starburst and interacting galaxies. Star cluster formation is 
an important mode of star formation, in particular in strong star formation 
regimes. I discuss the role of star clusters and their properties in helping 
us assess the question if star formation is a universal process allowing for
considerable scaling or if there's two different regimes for normal and violent
SF.}
\noindent
{\small¥{\it Keywords}: Stars: formation, Galaxies: evolution, formation, 
interactions, ISM, starburst, star clusters, globular clusters: general, 
open clusters and associations: general}

\section{Motivation}
Despite considerable efforts by many researchers over more than 30 yr, the 
question if {\bf S}tar {\bf F}ormation ({\bf SF}) is basically described by 
one universal process that can be scaled up and down considerably from lowest 
levels in dwarf and low surface brightness galaxies to extremely high levels 
observed in massive gas-rich interacting galaxies, ULIRGs, and SCUBA galaxies or 
if, on the other hand, there are two fundamentally different modes or processes 
of SF -- violent as opposed to normal SF -- still is one of the major unresolved 
issues in astrophysics. 

\section{Star Formation Rates: Measures and Regimes}
\subsection{Star Formation Rates: Measures and Limitations}
In nearby galaxies, spirals, irregulars and starburst galaxies, {\bf SF R}ates 
({\bf SFR}s) are conventionally derived from H$_{\alpha}$ or FIR luminosities 
using handy formulae like
$${\rm SFR ~[M_{\odot}/yr] ~=~ L(H_{\alpha}~/~1.26 \cdot 10^{41})~[erg/s]~~ 
(Kennicutt~1998)}$$
$${\rm SFR ~[M_{\odot}/yr] ~=~ L(FIR)~/~5.8 \cdot 10^9~[L_{\odot}]}$$
that are valid for a Salpeter IMF from 0.1 through 100 ${\rm M_{\odot}}$, 
for constant or slowly varying SFRs and for metallicities close to solar. 

Our evolutionary synthesis models GALEV for galaxies or star clusters of 
various metallicities do include gaseous emission in terms of lines and 
continuum on the basis of the time evolving and metallicity-dependent summed-up 
Lyman continuum photon rate (Smith et al. 2002, Schaerer \& de Koter 1997) from 
all the hot stars present in a cluster or a galaxy. While hydrogen emission 
line strengths are taken from photoionisation models, the line ratios of heavy 
element lines relative to ${\rm H_{\beta}}$ for low metallicities are taken 
from empirical compilations by Izotov et al. (1994, 1997) and 
Izotov \& Thuan (1998). Our models confirm 
the relations between H$_{\alpha}-$ and [OII]$-$ line luminosities in normal 
SF regimes and show their limitations: {\bf 1)} at 
significantly subsolar metallicities ${\rm Z \sim \frac{1}{20}~Z_{\odot}}$ SFRs 
estimated from ${\rm H_{\alpha}-}$ luminosities by the above formula are 
overestimated by factors $\leq 3$ for continuous SF regimes and by factors 
$\geq 3$ for starbursts. This is due to the fact that low-metallicity stars 
are hotter and have stronger ionising fluxes. {\bf 2)} In case of SFRs 
fluctuating on short timescales ($\leq 10^7$ yr) errors up to factors of 100 
can arise. For short-timescale fluctuations of the SFR, as e.g. in dwarf 
galaxies and their starbursts as well as in SF regions/complexes within larger 
galaxies, the delay of the ${\rm H_{\alpha}-}$ luminosity maximum with respect 
to the maximum of the SFR due to the fact that massive supergiants have even 
stronger ionising fluxes than their main sequence progenitors increases the 
errors in the SFRs estimated from ${\rm H_{\alpha}}$ beyond the metallicity 
effect discussed above. Fig. 1a shows the good agreement of our solar metallicity
const. SF model with Kennicutt's relation and the differences
arising for metallicities other than solar. Starbursts with durations from 
$10^5$ to $10^8$  
yr are put on top of the constant SFR models around ages of 10 Gyr. The rapid rise
of the SFR is not immediately reflected in a corresponding increase in 
${\rm H_{\alpha}-}$ luminosity: the ratio of log(${\rm
L(H_{\alpha})/SFR)}$ decreases by a factor up to 10 for the shortest burst 
before it strongly increases by a factor $\leq 60$ in the course of the burst 
and comes down to the pre-burst value after the burst (see Fig. 1b and Weilbacher \& FvA 2001 
for details). 

\begin{figure}[t]
\begin{center}
\psfig{figure=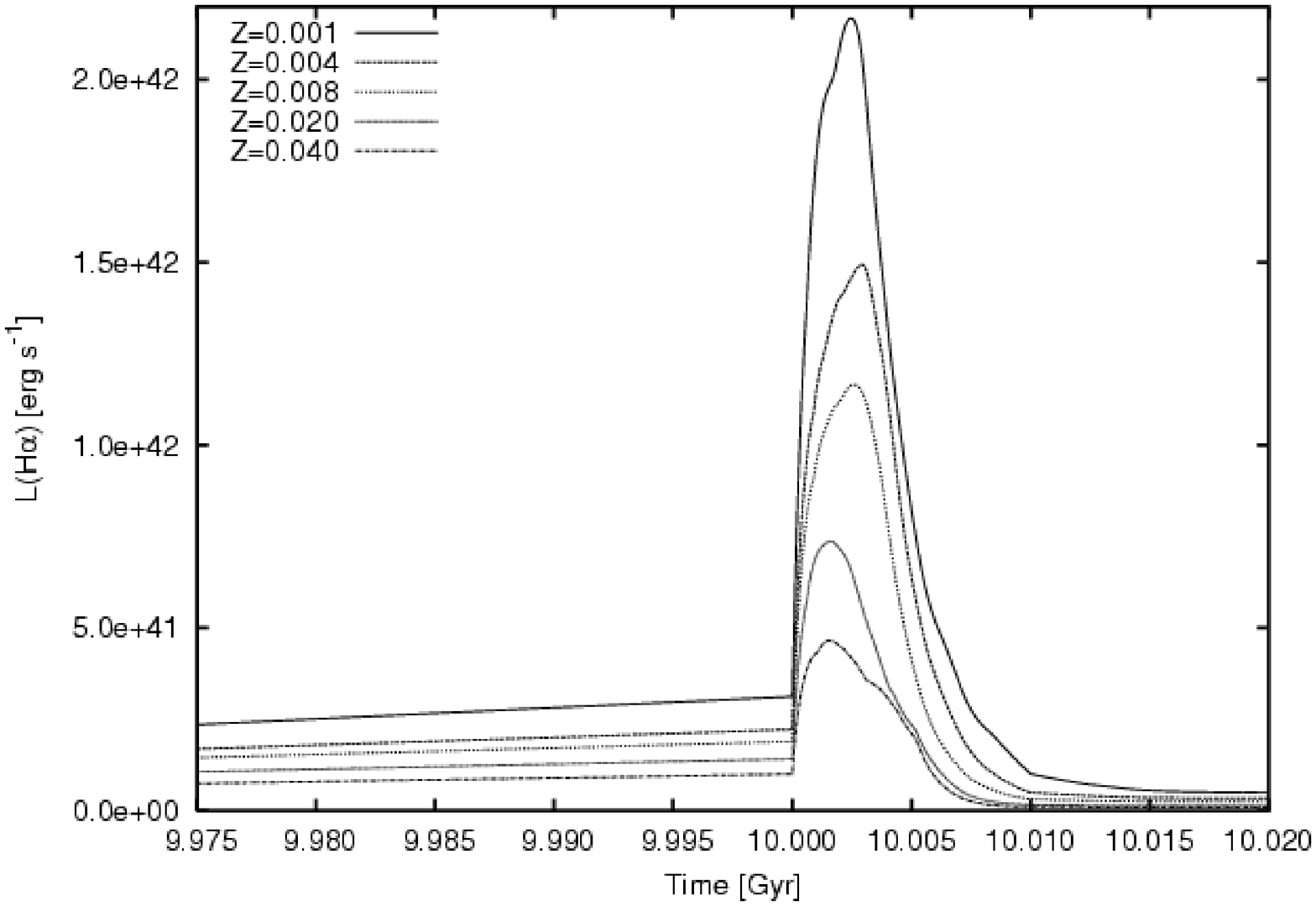,height=5cm}
\psfig{figure=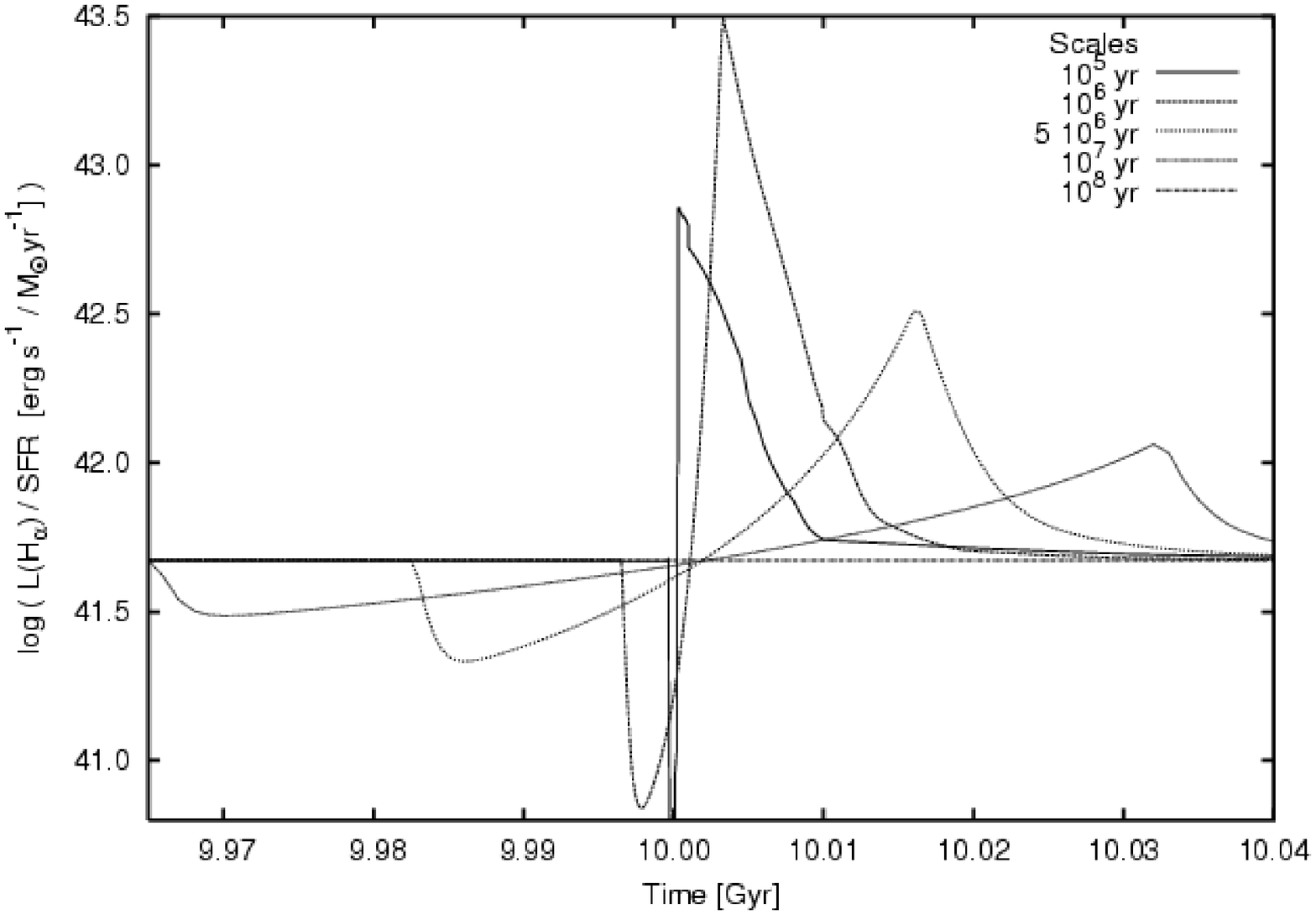,height=5cm}
\end{center}
\caption{Time evolution of L(H$_{\alpha}$) for const. SF plus a burst of $10^6$ yr duration starting after 10 Gyr in models with various metallicities (Fig. 1a) and of the ratio L(H$_{\alpha}$)/SFR for Gaussian shaped bursts of various durations for Z=0.001 (Fig. 1b).}
\end{figure}

SFRs of distant galaxies are often estimated from their [OII]3727$-$
luminosities. The metallicity dependence of the [OII]3727$-$line is twofold.
[OII] fluxes depend on the oxyen abundance and, hence, increase with increasing 
metallicity of the ionised gas, and on the strength of the ionising flux that
decreases with increasing metallicity. The combination of both effects 
accounts for 
a factor $\sim 2$ change in the transformation factor between ${\rm L([OII])}$ 
and SFR (see also Weilbacher \& FvA 2001). 

SFRs are particularly meaningful if expressed in relation to galaxy masses. 
In normal SF mode, spiral galaxies with typical masses of order 
${\rm 10^{10}~M_{\odot}}$ have global SFRs around ${\rm 1 - 3~M_{\odot}/yr}$. 
Irregular and dwarf irregular galaxies with masses in the range 
${\rm 10^6~to~10^9~M_{\odot} }$ have SFRs of order ${\rm 0.01 -
3~M_{\odot}/yr}$. Starbursts in dwarf galaxies, e.g. {\bf B}lue 
{\bf C}ompact {\bf D}warf 
{\bf G}alaxies ({\bf BCDG}s) feature SFRs of order ${\rm 0.1 -
10~M_{\odot}/yr}$.

\subsection{Burst Strengths}
Bursts strengths -- defined as the relative increase in stellar mass during 
the burst ${\rm b:=\Delta S_{burst} / S}$ -- in BCDGs have been shown 
by Kr\"uger, FvA \& Loose (1995) with evolutionary synthesis modelling 
compared to optical through NIR photometry to range from ${\rm b=0.001}$ to
${\rm b=0.05}$, 
and to decrease with increasing total mass of the galaxy, including M(HI), as
shown in Fig. 2, 
in agreement with expectations from stochastic self-propagating SF scenarios. 

\begin{figure}
\begin{center}
\psfig{figure=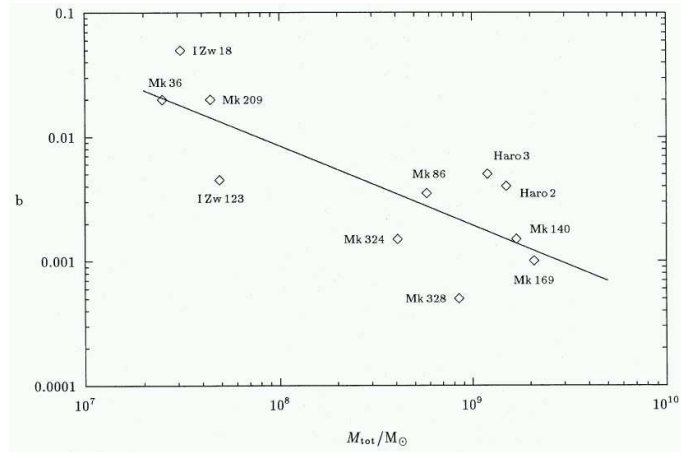,height=6cm}
\end{center}
\caption{Burst strengths as a function of total dynamical galaxy mass, including  HI, for the sample of BCDGs analysed by Kr\"uger {\it et al.} (1995).}
\end{figure}
 
All of the starbursting dwarf galaxies we have analysed so far, however, are
fairly isolated. Note that accurate burst 
strengths can only be given for systems at the end of the burst, i.e. for young 
post-starburst galaxies. As post-starbursts galaxies age, the precision to 
which burst strengths can be measured starts to decrease significantly as soon 
as the peak of the strong Balmer absorption line phase is over about 1 Gyr
after the burst. For galaxies with ongoing starbursts only lower limits to the 
burst strength can be estimated. 

Massive gas-rich interacting galaxies feature high and sometimes very high SFRs
of order $50,~100,$ and up to ${\rm 1000~M_{\odot}/yr}$ for {\rm L}uminous and
{\rm U}ltraluminous {\rm IR} {\rm G}alaxies, {\rm LIRG}s and {\rm ULIRG}s, in
their global or nuclear starbursts which typically last over a few $10^8$ yr. 
Evolutionary synthesis modelling of post-starbursts in massive gas-rich spiral --
spiral merger remnants like NGC 7252 have shown that these systems can also have
 tremendous bursts strengths that increase their stellar masses by $10-30$ and
possibly up to 50\% (FvA \& Gerhard 1994a, b). Starbursts in massive interacting galaxies hence are
completely off the burst strength -- galaxy mass relation for starbursts in
non-interacting dwarf galaxies. Whether this is due to the presence/absence of
an external trigger or due to the difference in gravitational potential and
dynamical timescale between
dwarf and giant galaxies is an open question still. Careful analyses of dwarf --
dwarf or giant -- dwarf galaxy mergers should tell the difference. 

\subsection{Star Formation Efficiencies}
Even under the most conservative assumptions -- that the
pre-merger spirals were drawn from the high end of their type-specific
luminosity function and were particularly gas-rich for their type -- the {\bf SF
E}fficiencies ({\bf SFE}s), defined in terms of SFE:$=$mass of stars/mass
of gas available, in these cases must have been extremely high. For the global
starburst happening slightly less than 1 Gyr ago during the merger of two
bright gas-rich spirals now called NGC 7252, the analysis of the deep Balmer
absorption line spectrum taken by F. Schweizer with our models indicated a
SFE $\geq 40$\% -- on a global scale (cf. FvA \& Gerhard 1994a, b). This should
be compared to the large-scale SFEs around $0.1-3$\% as determined for normal 
spiral and irregular galaxies and for starbursts in (isolated) dwarf galaxies.  

Small scale SFEs in Milky Way {\bf M}olecular {\bf C}louds ({\bf MC}s) can be 
defined as the mass ratio of the core mass of a MC to its total mass, 
SFE:${\rm =M(MC core)/M(MC)}$, and also have values in the range $0.1-3$\%. I.e., a very
small mass fraction of order $0.1$ to few \% of Galactic MCs has the high 
densities relevant for SF and makes up the cloud core. 

On scales of $10-300$ pc, ULIRGs that all have been shown to be advanced stages
of massive gas-rich galaxy mergers, feature SFEs in the range $30-100$\%. Their
extremely strong nuclear starbursts are heavily dust-obscured, emitting the bulk of their
bolometric luminosities at FIR wavelengths. 

\section{Molecular Clouds and Star Formation}
\subsection{Molecular Cloud Structure and Star Formation}
In the Milky Way and nearby galaxies, molecular clouds are observed using
sub-mm lines with different lines being tracers of molecular gas at different
densities. The most often observed CO(1$-$0) line traces gas at densities 
${\rm n \geq 100~cm^{-3}}$, the HCN(1$-$0) line traces gas at 
${\rm n \geq 30\,000~cm^{-3}}$, and the CS(1$-$0) line traces gas at 
${\rm n \geq 100\,000~cm^{-3}}$. Within the Milky Way and the nearest Local
Group galaxies detailed
observations of MC complexes in these different lines allow to assess their
internal structure. It is from this kind of observations that we know that
MCs typically have much more than 90\% of their mass in low density
envelopes as traced by CO and only a few \% in their high density cores as
traced by HCN or CS. I.e., 
on small scales for Milky Way MCs 
$${\rm L(HCN,~CS)/L(CO) \sim 0.1~-~3\%~ \sim ~M(MC~core)/M(MC)}$$

For galaxies beyond the Local Group, only integrated measures of
luminosities in these different lines are possible and 
allow to estimate integrated mass ratios of
molecular gas at various densities. On scales of 10$-$300 pc it has been shown
for ULIRGs that
$${\rm L(HCN,~CS)/L(CO) \sim 30~-~100\%~ \sim ~M(MC~core)/M(MC),}$$
suggesting that the MC structure in these massive gas-rich interacting galaxies
with their tremendous starbursts is drastically different from the MC structure
in the Milky Way or the Magellanic Clouds. The dynamical mass in the central
regions of ULIRGs is dominated by the mass of molecular gas at densities of MC
cores. This is, in fact, predicted by hydrodynamical simulations of gas-rich 
spiral -- spiral mergers and necessary if these mergers are to result in
elliptical galaxies. The molecular gas densities in the centers of ULIRGs are
similar to the central stellar densities in giant ellipticals, their SFRs
high enough to transform the gas into stars within the typical duration of a 
ULIRG phase ($1-4 \cdot 10^8$ yr). These apparent drastic 
differences in MC structure immediately raise the question if the SF process in
massive gas-rich galaxy mergers, in particular in those going through a LIRG or 
ULIRG phase, can be the same as for the normal SF mode of undisturbed galaxies
or for the mini-starbursts in BCDGs -- as they appear in comparison to those in
LIRGs or ULIRGs? A scenario where increased rates of cloud -- cloud collisions
are at the origin of the enhanced SF is hard to imagine in view of the fact that
the core of Arp 220 on a scale of several 10 to 100 pc resembles {\bf one}
ultra-giant MC core of order ${\rm 10^{10}M_{\odot}}$ (cf. FvA 1994).

Observationally, a very tight correlation between global SFRs measured from FIR
luminosities and the total mass in MC cores measured in terms of
HCN$-$luminosities is found to hold for normal spirals as well as for the
most extreme LIRGs and ULIRGs, i.e. over 4 orders of magnitude in terms of both 
total MC core masses and SFRs (Solomon {\it et al.} 1992, Gao \& Solomon 2004).
At the same time, the ratio between SFR as measured from L$_{\rm FIR}$ and total
MC core mass as measured by L$_{\rm HCN}$ is roughly constant for all SFing
galaxies from spirals through ULIRGs (Gao \& Solomon 2004), indicating that the 
SF efficiency when
referred not to HI but to the amount of dense molecular gas as traced by HCN or
CS is constant over all the dynamical range. 

For all galaxies (BCDGs ... Irrs ... spirals ... ULIRGs), SF efficiencies
quantitatively correspond to the ratio between the integrated mass of MC cores
and the total mass of molecular gas. For BCDGs ... spirals the total mass of
molecular gas exceeds that of the MC cores by a factor $\geq 100$, for ULIRGs
both quantities are comparable. Hence, 
SFE $\sim$ M(MC core)/M(MC) or L(FIR) $\sim$ L(HCN, CS)/L(CO).

The widely used Schmidt (1959) law relates the SFR density to the neutral or
molecular gas density to a power n with n$\sim 1$ for spirals, Irrs, and BCDGs,
and n$\sim 2$ for ULIRGs and holds over 5 orders in gas surface density and 6
orders in SFR density. 

When expressed in terms of high density gas traced by HCN or CS, the Schmidt law
takes the form 
$${\rm SFR~density \sim (gas(HCN,~CS)~density)^n}$$
with ${\rm n = 1}$ for {\bf all} galaxies (spirals, . . ., ULIRGs) and all SF
regimes, as also shown by Gao \& Solomon (2004).

In the course of mergers among gas-rich galaxies, hydrodynamic models (SPH) as well as
sticky particle codes predict strongly enhanced collision rates among MCs that
push up their SFRs. Models also predict that shock compression of MCs should
significantly raise SF efficiencies to values ${\rm SFE \leq 0.75~-~0.9}$,
already for small overpressures in the intercloud medium (Jog \& Das 1992,
1996). Strong burst SFRs require not only pre-existing MCs to be efficiently
transformed into stars but also the fast transformation of HI into molecular
gas. 
McKee \& Ostriker (1977) have shown that shocks are very efficient in
promoting the transformation of HI, leaving the ISM behind strong shocks almost
fully molecular. 

It hence appears that once gas is compressed to MC core densities, it is with
almost 100 \% efficiency transformed into stars. The process that determines the
SF timescale and the SF efficiency seems to be the compression of gas to these
high densities. And this process, in turn, is apparently slow and has low
efficiency in non-interacting spirals, irregulars and even starbursting dwarfs,
while fast and very efficient in massive gas-rich interacting galaxies.

\subsection{Molecular Cloud Mass Spectra}
In spiral and irregular galaxies and normal SF mode MCs, their cores, and 
ultimately even the star
clusters that form from them, all feature similar mass spectra that are power
laws with index ${\rm m \sim -1.7 \dots -2}$. 

Largely unexplored at present are the mass spectra of MCs and MC cores in
strongly interacting galaxies due to the large distance of those systems. 
A first attempt in this direction is presented by Wilson {\it et al.} (2003) for
the Antennae galaxy pair NGC 4038/39 at a distance of ~15 Mpc, an ongoing merger
of two Sc-type spirals as estimated from the HI-richness of their long tidal
tails. NGC 4038/39 is a LIRG with the most vigorous SF going on in the overlap
region between the two disks and huge amounts of star cluster formation. 
Ground- and space-based observations
over a large wavelength range as well as extensive dynamical modelling is
available for the Antennae galaxies, the youngest system in Toomre \& Toomre's
(1972) age sequence of interacting galaxies. 
Wilson {\it et al.} find the mass spectrum of MCs in the Antennae to
obey a power-law with m in the range $-1.2 \dots -1.6$ in the accessible mass
range from $10^7$ to ${\rm 10^9~M_{\odot}}$. The mass range below 
${\rm 10^7~M_{\odot}}$ as well as the mass spectrum of MC cores remain
inaccessible to present-day instrumentation. Although this slope is slightly
flatter than for MC mass spectra in non-interacting galaxies, it is not clear
yet, if the MC mass spectrum in the Antennae is really enhanced in massive MCs 
due to the high ambient pressure as could be expected from the above-quoted
models. 

\section{Star Cluster Formation}
Star cluster formation is an important mode of SF, in particular in starbursts.
$\sim 20$ \% of the UV luminosity of starburst galaxies is accounted for by 
Star Clusters, and the contribution of star clusters to the total UV
luminosity seems to increase with increasing UV surface brightness (Meurer {\it
et al.} 1995). 

Star clusters observed with HST in a large number of interacting and merging
galaxies and young merger remnants seem to span the full range from low mass
clusters ($\sim 10^3~M_{\odot}$) through high and very high mass clusters 
($\geq 10^7~M_{\odot}$), from weakly bound, short-lived clusters similar to the
open clusters in nearby galaxies all through strongly bound and long-lived
clusters analoguous to Globular Clusters. 

It has been predicted by hydrodynamical cluster formation models that the
formation of strongly bound and hence long-lived clusters requires very 
high SF efficiencies ${\rm SFE \geq
20}$ \% (Brown {\it et al.} 1995), and is therefore generally not 
possible during normal SF in spiral or
irregular galaxies, nor in the mini-starbursts in BCDGs. 

The very existence of a large number of massive compact star clusters in the
relatively old spiral -- spiral merger remnant NGC 7252, in which 
a very strong burst ended more than  
Myr ago, proves that these clusters must be very strongly bound -- like Globular
Clusters -- as they survived for that span of time in an environment where 
violent relaxation has been strong enough to  
transform the two spiral disks into an elliptical-like object with an ${\rm
r^{1/4}}$ light profile (Schweizer 2002).

In the Antennae NGC 4038/39 we have analysed the 550 star clusters that have been detected
in V and I with HST WFPC1 by Whitmore \& Schweizer (1995) with our GALEV models
and derived ages from their V-I colors under the assumption that they have
around half-solar metallicity -- as expected if they form from the ISM of Sc
spirals and confirmed by spectroscopy of the brightest of them by Whitmore {\it
et al.} (1999). We found
480 of them to have ages $\leq 4 \cdot 10^8$ yr and 70 to be fiducially old
Globular Clusters inherited from the progenitor galaxies (FvA 1998). We followed
their evolution with our GALEV evolutionary synthesis models and showed that --
provided they would all survive -- they would develop a color distribution 
with the same width but somewhat redder, due to 
their enhanced metallicity, as those of metal-poor GCs and a Gaussian shape {\bf L}uminosity {\bf F}unction
({\bf LF}) typical of old GC systems despite the fact that their observed LF is
a power law. It is the age spread among the young star clusters, that is
comparable to their age, in conjunction with the rapid luminosity evolution
during these young ages and with the observational completeness limit that
causes this apparent distortion in the LF. In FvA (1999) we derived masses from
ages and model M/L-ratios for all the clusters and found the {\bf M}ass {\bf F}unction
({\bf MF}) of the {\bf young} star cluster system to be a Gaussian with ${\rm 
\langle log~M \rangle=5.6}$ and ${\rm \sigma =0.46}$ very similar to that of GCs
in the Milky Way and M31 with ${\rm 
\langle log~M \rangle=5.47}$, ${\rm \sigma =0.5}$ and ${\rm 
\langle log~M \rangle=5.53}$, ${\rm \sigma =0.43}$, respectively (cf. Ashman
{\it et al.} 1995). 

\begin{figure}[t]
\begin{center}
\psfig{figure=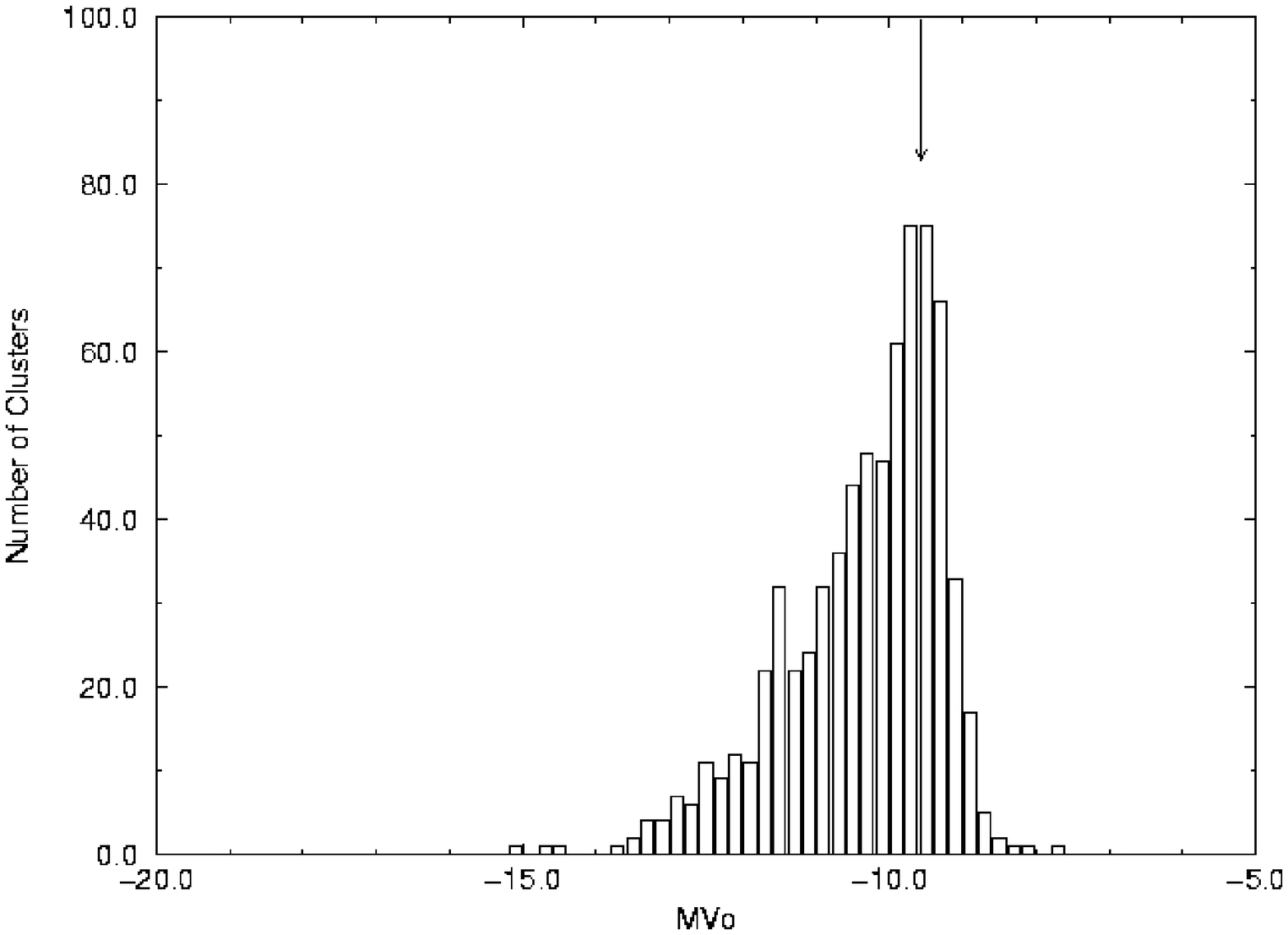,height=6cm}
\psfig{figure=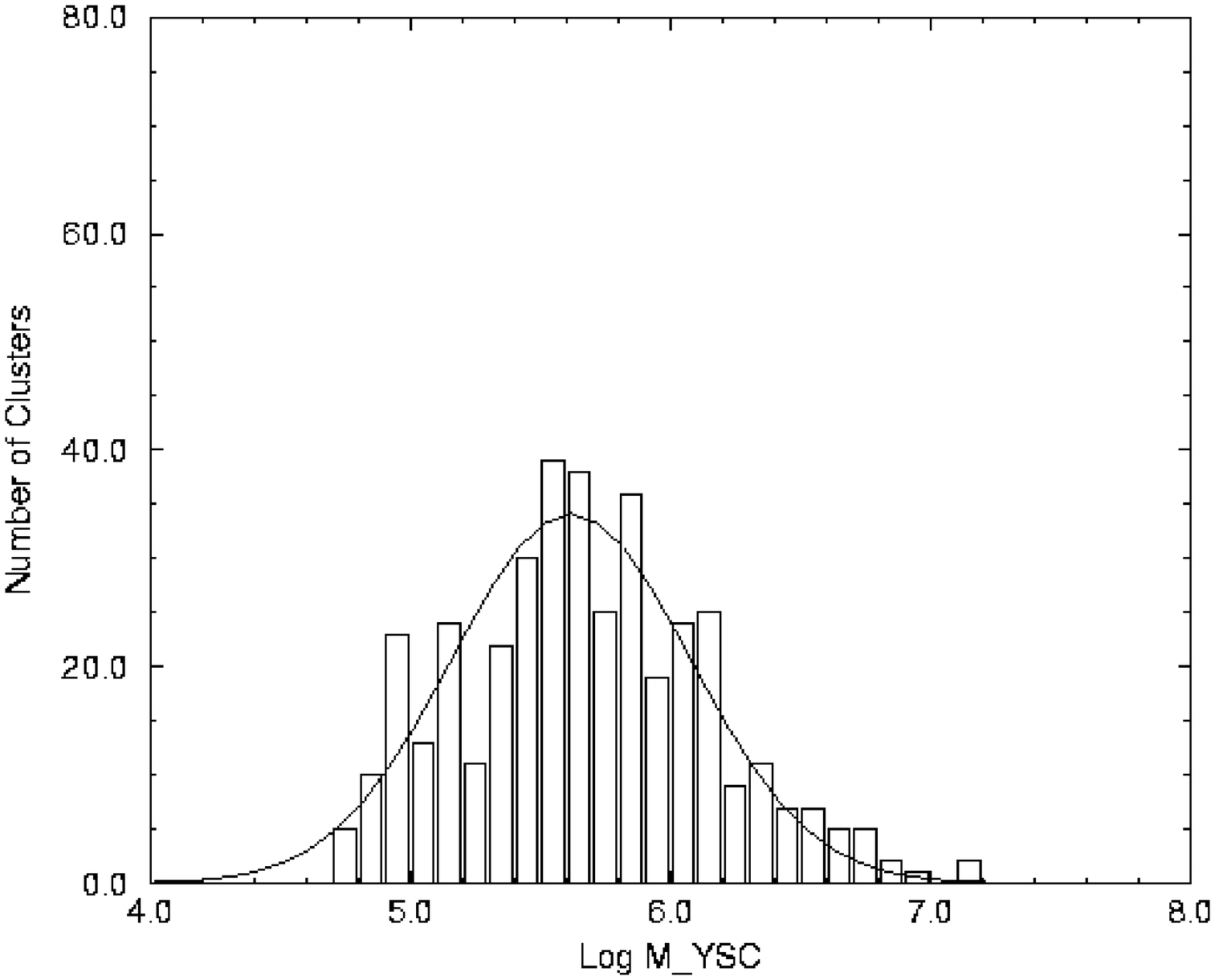,height=6cm}
\end{center}
\caption{Observed LF of star clusters in NGC 4038/39 (Fig. 3a). The arrow indicates the observational completeness limit. 
Mass Function derived for the {\bf young} clusters with a Gaussian fit as described in the text (Fig. 3b).}
\end{figure}

The major drawback in our analysis was our
assumption of a uniform reddening for all young clusters lack of more detailed
information about individual clusters. Zhang \& Fall (1999) used reddening-free
Q-parameters for their analysis of the same data and found a power-law mass
function. The major drawback in their analysis was that they had to exclude an 
important fraction of clusters for which the Q-parameters did not yield an
unambiguous age. Excluding this age group of clusters in our models also leads
to a power-law MF. Hence, till today, the MF of young star clusters forming in
merger-induced starburst is controversial. A multi-wavelength analysis should
allow to independently determine metallicities, ages, extinction values, and
masses of all the young star clusters provided accurate photometry in at least 4
reasonably spaced passbands is available, as shown by Anders {\it et al.}
(2004a), and is currently underway using HST WFPC2, NICMOS and VLT data provided
by our ASTROVIRTEL project (PI R. de Grijs). 

The question is if and to what percentage the young star clusters copiously formed in
galaxy mergers are open clusters or GCs and if they split into these two 
distinct classes of
objects or if there is a continuum extending from losely bound and low-mass open
clusters to strongly bound and high-mass GC. Key issues for this question are 
their mass range, their MF and
their compactness. Size determinations for young star clusters require a careful
analysis: small clusters are barely resolved even in the closest interacting
systems, the galaxy background is bright and varies on small scales, and some 
clusters do not (yet?) seem to be tidally truncated, i.e.
cannot be described by King models. The degree of internal binding, i.e. the 
ratio between mass and radius, however, is a key parameter for survival or
destruction of a cluster in the violently changing environment of the merging
and relaxing galaxy. 

\subsection{Globular Cluster Formation}
GC formation requires extremely high SF efficiencies. It happened in the Early
Universe and it apparently happens today in the strong starbursts accompanying
the mergers of massive gas-rich galaxies. If it also happens in
non-interacting massive starburst galaxies or in dwarf galaxy starbursts is an
open question. 

Our investigation of star clusters in the dwarf starburst galaxy NGC 1569, that was
known before to host 3 super star clusters, revealed $\sim 160$ young star
clusters with good photometry in many bands in our ASTROVIRTEL data base. Analysis of their {\bf S}pectral {\bf E}nergy 
{\bf D}istributions ({\bf SED}s) in comparison with a large grid of GALEV models
for star clusters with various metallicities and dust extinctions by means of 
a dedicated SED Analysis Tool yielded individual clusters ages -- all $\leq 24$ 
Myr --, metallicities, extinction values, and masses. As seen in Fig. 4, masses
of all but 3 of these clusters turn out to be lower and most in fact much lower
-- of order ${\rm 10^3-10^4~M_{\odot}}$ -- than those of GCs in the Milky Way
despite their high luminosities that are due to their very young ages 
(Anders {\it et al.} 2004b).

\begin{figure}[t]
\begin{center}
\psfig{figure=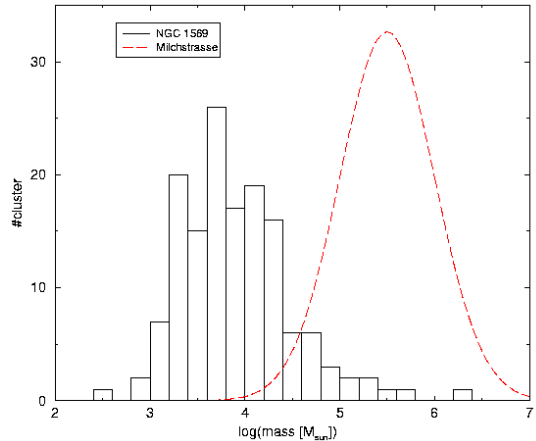,height=6cm}
\end{center}
\caption{Mass Function of young star clusters in the dwarf starburst galaxy NGC
1569 as compared to the MF of Milky Way GCs.}
\end{figure}

Hence, with maybe 2 or 3 exceptions -- depending on
a careful determination of their sizes --, the rich bright young cluster 
population in the dwarf starburst galaxy NGC 1569 does not seem to comprise any
young GCs, most of its low-mass clusters will probably not survive the next
$1-2$ Gyr. This raises the question why GCs do not form in dwarf galaxy
starbursts. Why are SF efficiencies low in dwarf starbursts as already found for
BCDGs many years ago? Because of the short dynamical timescales or the shallower
potential in dwarf galaxies or because of a lack of ambient pressure in these
non-interacting galaxies as
compared to massive interacting galaxies? An answer should be provided by
careful analyses of the starburst and star cluster properties in dwarf -- dwarf
galaxy mergers.

\subsection{Star Cluster vs. Field Star Formation}
An intriguing example of episodes with and without cluster formation is provided
by the LMC. It shows a clear gap in terms of star cluster ages 
(Rich {\it et al.} 2001)
with no clusters in the age range from 4 -- 10 Gyr. This gap, however, is
not seen in field star ages and the metallicity apparently has also increased
continuously over the cluster age gap. Star cluster formation epochs coincided
with epochs of enhanced field star formation, probably associated with close
encounters between the LMC and the Milky Way. 

\section{Conclusions and Open Questions}
I have shown that global galaxy-wide SFRs span a huge range, even in relation 
to galaxy mass, from normal low-level SF in undisturbed disk galaxies to the
extremely high SFRs in massive gas-rich interacting galaxies, ULIRGs, and
SCUBA galaxies. 

I cautioned that SFR estimates from H$_{\alpha}-$ or O[II]$-$ luminosities are
only valid for metallicities close to solar and for SFR fluctuations on
timescales $\geq 10^8$ yr, hence not for dwarf galaxy starbursts, nor for SFing
regions on subgalactic scales. 

Concerning SF efficiencies, there is a clear dichotomy between normal galaxies
and dwarf galaxy starbursts on the one hand and starbursts in massive
interacting gas-rich galaxies on the other hand, with SFEs differing by factors
10$-$100 between them. It apparently originates in a similar dichotomy for the
integrated mass ratio between molecular gas at low densities as traced by CO and the high
density molecular gas of MC cores as traced by HCN or CS with the ratio ${\rm
M(MC~core)/M(MC)}$ differing by the same factor 10$-$100. The key process
determining the SF efficiency seems to be compression of molecular gas to MC
core densities. Once this is accomplished, the high density MC core material is
transformed into stars with very high efficiency -- in fact with the {\bf same}
efficiency in normal, starburst, and ULIRG galaxies. 

The causes of these differences are not clearly identified yet. They could be
differences in the dynamical timescales, in the depth of the potential or the
dynamics of a merger. Detailed investigations into the starburst and its star and
cluster formation in a dwarf -- dwarf galaxy merger should tell.

While it will not be possible to resolve the masses of MCs and MC cores down to
interesting values before ALMA -- not even for the closest interacting galaxies, 
the comparison of integrated luminosities in lines tracing molecular gas at
various densities should already yield interesting clues to the molecular cloud
structure in various kinds of starbursts. 

Star cluster formation is an important and sometimes dominant mode of SF. It is
not clear yet if the mass ratio between SF going into field stars and SF going 
into star cluster formation -- and, in particular, into the formation of compact
massive long-lived GCs -- scales with the strength of SF or burst, or with the
SF efficiency. A comparative investigation of integrated starburst properties
and those of the young star cluster populations should help. 

A third dichotomy, probably related to the other two, was found concerning GC 
formation. While GC formation apparently is possible and wide-spread in high SF efficiency
situations as in the Early Universe or in massive gas-rich spiral -- spiral
mergers, it does not seem to be possible, or at least not frequent, in 
isolated dwarf galaxy starbursts.

The age and metallicity distributions of GC (sub-)populations contain valuable
information about the violent (star) formation histories of their parent
galaxies and can reasonably be disentangled by means of multi-wavelength SED
analyses. 

\section*{Acknowledgments} I thank the organisers and the DFG (Fr 916/11-1) 
for partial travel support and gratefully acknowledge very efficient support 
from the 
ASTROVIRTEL team (project funded by the European Comm. under FP5
HPRI-CT-1999-00081).

\end{document}